\documentclass[aps,prc,twocolumn,superscriptaddress,nofootinbib,tightenlines,
preprintnumbers,showkeys,floatfix]{revtex4-2}

\usepackage[utf8]{inputenc}
\usepackage[english]{babel}
\usepackage{amssymb,amsthm,amsmath,amstext,amsbsy,amsopn,mathrsfs}
\usepackage{bbm}
\usepackage{nicefrac}
\usepackage{slashed}
\usepackage{graphicx}
\usepackage{hyperref}
\usepackage{leftidx}
\usepackage{environ}
\usepackage{mathtools}
\usepackage{xspace}
\usepackage{array}
\usepackage{xcolor}
\usepackage{suffix}
\usepackage{booktabs,colortbl}
\usepackage{pgfplots}
    \pgfplotsset{compat=1.18}
    \usepgfplotslibrary{groupplots}
    \usetikzlibrary{arrows.meta}
    \usetikzlibrary{external}
\usepackage{isotope}
\usepackage{braket}
\usepackage{xfrac}
\usepackage{gensymb}

\newcommand{\ie}{\textit{i.e.}\xspace}

\newcommand{\apriori}{\textit{a priori}\xspace}
\WithSuffix\newcommand\apriori*{\textit{a-priori}\xspace}

\newcommand{\ii}{\mathrm{i}}
\newcommand{\ee}{\mathrm{e}}

\newcommand{\Rp}{\mathrm{Re}}
\newcommand{\Ip}{\mathrm{Im}}

\newcommand{\abs}[1]{\left|#1\right|}

\newcommand*\rvec[1]%
{\ensuremath{\overset{\smash{\raisebox{-1.5pt}{\tiny$\rightarrow$}}}{#1}}}
\newcommand*\lvec[1]%
{\ensuremath{\overset{\smash{\raisebox{-1.5pt}{\tiny$\leftarrow$}}}{#1}}}

\NewEnviron{subalign}[1][]{%
\begin{subequations}\begin{align}
  \BODY
\end{align}\label{#1}\end{subequations}
}

\NewEnviron{spliteq}{%
\begin{equation}\begin{split}
  \BODY
\end{split}\end{equation}
}

\NewEnviron{mleq}{%
\begin{equation}
  \BODY
\end{equation}
}

\renewcommand{\vec}[1]{\mathbf{#1}}

\newcommand{\cphase}[1]{\ee^{{#1}\ii\phi}}

\begin{document}

\title{Towards scalable bound-to-resonance extrapolations for few- and many-body systems}

\author{Nuwan Yapa}
\email{nyapa@fsu.edu}
\affiliation{Department of Physics, Florida State University,
Tallahassee, Florida 32306, USA}

\author{Sebastian K\"onig}
\email{skoenig@ncsu.edu}
\affiliation{Department of Physics, North Carolina State University,
Raleigh, North Carolina 27695, USA}

\author{K\'evin Fossez}
\email{kfossez@fsu.edu}
\affiliation{Department of Physics, Florida State University,
Tallahassee, Florida 32306, USA}
\affiliation{Physics Division, Argonne National Laboratory, Lemont,
Illinois 60439, USA}

\begin{abstract}
In open quantum many-body systems, the theoretical description of resonant states of many particles strongly coupled to the continuum can be challenging. 
Such states are commonplace in, for example, exotic nuclei and hadrons, and can reveal important information about the underlying forces at play in these systems.
In this work, we demonstrate that the complex-augmented eigenvector continuation (CA-EC) method, originally formulated for the two-body problem with uniform complex scaling, can reliably perform bound-to-resonance extrapolations for genuine three-body resonances having no bound subsystems. 
We first establish that three-body bound-to-resonance extrapolations are possible by benchmarking different few-body approaches, and we provide arguments to explain how the extrapolation works in the many-body case.
We furthermore pave the way towards scalable resonance extrapolations in many-body systems by showing that the CA-EC method also works in the Berggren basis, studying a realistic application using the Gamow shell model.
\end{abstract}

\maketitle

\section{Introduction}
\label{sec:Introduction}

Broad many-body resonances represent a new frontier in quantum physics. 
For instance, in hadron physics, searches for exotic structures allowed by quantum chromodynamics, such as tetra- and pentaquarks~\cite{olsen18_3060,karliner18_325}, glueballs exclusively made of gluons~\cite{ablikim24_3059}, or structures involving heavy quarks, often involve arguments about the nature of the resonant peak observed. 
Similarly, in low-energy nuclear physics, the exploration of the limits of nuclear stability leads to the production of exotic nuclei.
The exact nature of these states can be controversial, as is the case, for example, for 
\isotope[9,10]{He} states~\cite{kalanee13_1909,votaw20_2353,vorabbi18_1977,fossez18_2171}
or the four-neutron system~\cite{kisamori16_1463,duer22_2494,hiyama16_1624,fossez17_1916,deltuva18_2079,higgins20_2495,li19_2634,lazauskas23_2744}.
Fundamentally, the goal is to establish the existence of a many-body quasistationary state in systems in which several particles are strongly coupled to the continuum of scattering states. 
In other words, we must be able to tell whether or not, at one point, all the particles constituting the system interacted with each other.
This is the reason why reliable theoretical approaches in this regime are needed.

In the past few decades, many-body methods formulated in the quasistationary formalism have emerged as a promising avenue towards scalable calculations of resonances~\cite{myo20_2674,michel21_b260}. 
However, these tend to struggle with broad many-body resonances due to the increasing computational cost and decreasing stability of calculations, as the number of particles increases. 
More recently, the introduction of the eigenvector continuation (EC) technique~\cite{frame18_2071}, a form of a reduced-basis method, opened many new possibilities (see Ref.~\cite{duguet24_3061} for a recent review). 

Eigenvector continuation constructs a low-dimensional representation of a given Hamiltonian, controlled by a set of parameters, using eigenstates of the Hamiltonian 
in the original large Hilbert space for various sets of parameters. 
It was shown that this low-dimensional representation can then provide accurate extrapolated eigenstates in regions of the parameter space where direct high-resolution calculations would be difficult or impossible.

In Ref.~\cite{yapa_eigenvector_2023}, we proposed a generalization of the EC method in the quasistationary formalism using the uniform complex-scaling method, called \emph{complex-augmented eigenvector continuation} (CA-EC), and showed that CA-EC enables accurate bound-to-resonance (B2R) extrapolations.
That is, if the interaction between the particles depends on a set of parameters in such a way that in a given region of the parameter space the system is bound, and in another region the system supports a resonance, then CA-EC makes it possible to efficiently predict the resonance energy using a reduced basis constructed in the bound-state regime.
In this work, we consider the application of CA-EC to few-body systems.
Specifically, we demonstrate that accurate B2R extrapolations of genuine three-body resonances are possible using the CA-EC method implemented with different numerical approaches, namely a configuration interaction (CI) method based on harmonic oscillator states and a discrete variable representation (DVR) in finite volume.
In addition, we show that a rigorous path towards scalable CA-EC calculations exists by using the Berggren basis, which is important for future many-body applications of the technique.

While going from two- to three-body systems might seem trivial, this actually represents a significant step. 
Indeed, three-body systems can exhibit various decay dynamics, such as partitioning into two stable subsystems (two-body decay), or into a single particle plus a resonant two-body subsystem with a long enough delay between the first and second decay so that the three-body decay can be described as sequential (``factorized'') two-body decays. 
The complexity of three-body decay is elegantly illustrated, for instance, in Fig.~25 of Ref.~\cite{pfutzner12_1169} in the context of two-proton emission. 
We must thus ensure that the model used to test CA-EC can produce genuine three-body resonances.
In addition, representing three-body outgoing asymptotics can be significantly more difficult than representing two-body asymptotics, and thus it must be checked that CA-EC still performs well in this case.
We show that this is indeed the case and provide an argument as to why CA-EC works as well as it does that is more general than what was previously discussed in Ref.~\cite{yapa_eigenvector_2023}. 
This is critical to establish the scalability and generality of the method.
We note that EC applied to non-Hermitian systems was also recently studied in Refs.~\cite{Zhang:2024ril,Liu:2024pqp,Zhang:2024gac} with a different focus than that of the present work.

Our paper is organized as follows. In Sec.~\ref{sec:Formalism}, we present the formalism used in this work, and notably the CA-EC method in Sec.~\ref{sec:CA-EC}. 
Results for different examples are presented in Sec.~\ref{sec:Results}. 
Finally, our conclusion and outlook are given in Sec.~\ref{sec:Conclusion}.

\section{Formalism}
\label{sec:Formalism}

In this section, after providing a brief overview of the coordinate systems used in this work (Sec.~\ref{sec:coordsyst}), we introduce the uniform complex-scaling method and the Berggren basis expansion technique (Sec.~\ref{sec:nonHermitian}) that we use to obtain Gamow states.
Then, we present the generalization of the CA-EC method with complex scaling for few and many-body resonances (Sec.~\ref{sec:CA-EC}) and its counterpart formulated with the Berggren basis (Sec.~\ref{Sec:result2bGSM}).

\subsection{Coordinate systems}
\label{sec:coordsyst}
In Sec.~\ref{sec:Results}, we will be using two different techniques to implement complex scaling, namely a harmonic oscillator (HO) basis as a simple case of a CI calculation that also goes by the name No-Core Shell Model (NCSM)~\cite{barrett13_688} in nuclear physics, as well as a discrete variable representation (DVR) in finite volume (FV).
These two formalisms use different choices for the coordinates that describe the few-body system, which we briefly introduce here.

For a system of $n$ particles all with equal mass $m$ we denote the single-particle positions as $\vec{x}_i$ for $i=1,\ldots,n$, and their momenta by $\vec{k}_i$.
Jacobi coordinates, which we denote as $\vec{r}_i$ in the following, can be written as a transformation
\begin{equation}
 \vec{r}_i = \sum_j U_{ij} \vec{x}_j \,,
\label{eq:Jacobi-u}
\end{equation}
where $U$ is a unitary matrix the detailed form of which can be found, for example,
in Ref.~\cite{Varga:1998}.
The transformation includes a coordinate for the overall center-of-mass position, which can in general be dropped because we are only interested in the intrinsic properties of states.
For the special case $n=3$, there are two relevant Jacobi coordinates\footnote{Note that there are equivalent choices arising from permuting the single-particle coordinates.}
\begin{align}
  \vec{r}_1 &= \vec{x}_2 - \vec{x}_1 \, , \\
  \vec{r}_2 &= \vec{x}_3 - \frac12\left(\vec{x}_1 + \vec{x}_2\right) \,.
\end{align}
We denote the momenta conjugate to the $\vec{r}_i$ as $\vec{p}_i$.

Jacobi coordinates have the feature that the kinetic (free) part of the Hamiltonian, which we denote by $H_0$, can be written purely in terms of second derivatives with respect to the individual $\vec{r}_i$ (or, equivalently, in terms of squares of the $\vec{p}_i$), which is 
often desirable for \textit{ab initio} calculations.
For other approaches, such as the FV-DVR that we use in Sec.~\ref{sec:Results}, it is more convenient to work in simple relative coordinates, which can be written as
\begin{equation}
 \tilde{\vec{r}}_i = \begin{cases}
  \vec{x}_i - \vec{x}_n & \text{for}\ 1 \leq i < n \,, \\
  \frac{1}{n} \sum_{j=1}^n \vec{x}_j & \text{for}\  i=n \,.
 \end{cases}
\label{eq:simple_relative_coordinates}
\end{equation}
This definition expresses all positions relative to the position of the ``last'' particle, and it can again be written as a unitary transformation $\tilde{\vec{r}}_i = \sum_j U_{ij} \vec{x}_j$.
In this coordinate system, the free Hamiltonian involves mixed terms, \textit{viz.},
\begin{equation}
    H_0 = \frac{1}{m} \sum_{i=1}^{n-1} \sum_{j=1}^{i} \tilde{\vec{p}}_i \cdot \tilde{\vec{p}}_j \,,
\label{eq:H0_in_SRC}
\end{equation}
where $\tilde{\vec{p}}_i$ is the momentum conjugate to $\tilde{\vec{r}}_i$, and relative separations between any pair of particles are easy to compute.

\subsection{Non-Hermitian formalism}
\label{sec:nonHermitian}

In this work, we describe resonances in the quasistationary formalism, \ie, as eigenstates of the time-independent Schr\"odinger equation with outgoing boundary conditions. 
Such states are known as Gamow or Siegert states and have complex energies~\cite{gamow28_500,siegert39_132}. 
In practice, two efficient ways to describe such states are the uniform complex-scaling method~\cite{reinhardt82_708,Moiseyev:1998aa} and the Berggren basis expansion~\cite{Berggren:1968zz,Berggren:1993zz}. 
Both techniques lead to complex-symmetric (non-Hermitian) Hamiltonian matrices.
Below, we briefly outline how these methods work.

\subsubsection{Uniform complex-scaling method}
\label{Sec:cxscaling}

Complex scaling, in its simplest form known as ``uniform complex scaling,'' amounts to multiplying the coordinate vectors by a complex phase.
If for the moment we consider a two-body system with relative coordinate $\vec{r}$
and momentum $\vec{p}$ (note that for $n=2$ the different coordinate choices discussed in Sec.~\ref{sec:coordsyst} coincide), we have
\begin{equation}
    \vec{r} \to \vec{r} \cphase{} \,,
\label{eq:CSM-for-r}
\end{equation}
and we see that the momentum should be scaled by an inverse factor,
\begin{equation}
    \vec{p} \to \vec{p} \cphase{-} \,,
\label{eq:CSM-for-p}
\end{equation}
in order to preserve the canonical commutation relation
\begin{equation}
    \left[ \vec{r} \cphase{}, \vec{p} \cphase{-} \right] = \left[ \vec{r}, \vec{p} \right] =  \ii \, .
\end{equation}

To implement complex scaling, we note that the momenta $\vec p$ in the free Hamiltonian
would be multiplied by scalar factors of $\cphase{-}$,
which propagates trivially to the front.
Therefore, under complex scaling, the Hamiltonian operator is transformed into
\begin{equation}
    H \rightarrow H_\phi = \frac{\cphase{-2} {\vec p}^2}{2 \mu} + V(\vec r \cphase{}) \,.
\end{equation}
For the potential part, this requires analytically continuing the interaction to evaluate it for complex arguments.

For the few and many-body cases, one can start by applying the complex scaling to the underlying single-particle coordinates and then, for both choices of relative 
coordinates introduced in Sec.~\ref{sec:coordsyst}, the overall effect is the same because they arise from the single-particle coordinates as linear transformations.
Thus, for Jacobi coordinates, we have
\begin{equation}
    \vec{r}_i \to \vec{r}_i \cphase{} \,, \quad
    \vec{p}_i \to \vec{p}_i \cphase{-} \quad \text{for all} \, i \, ,
\end{equation}
and likewise for simple relative coordinates.
Therefore, for any choice of coordinates, the free Hamiltonian $H_0$ will pick up a scalar factor $\cphase{-2}$, while the potential energy operator 
needs to be evaluated with complex scaled arguments.
For example, if $V$ is a local central two-body potential and $\vec{r}_{ij}$ is the relative distance between particles $i$ and $j$, then complex scaling is implemented as
\begin{equation}
    V(\vec{r}_{ij}) \to V(\vec{r}_{ij} \cphase{})
    \equiv V(|\vec{r}_{ij}|\cphase{}) \,.
\end{equation}
We note in particular that if the original potential depends only
on the magnitude of the separation between the particles, then the phase factor needs to be included \emph{after} evaluating the Euclidian norm of the separation vector, as indicated in the equation above.
This is discussed in more detail in Ref.~\cite{yu_complex_2024}, where the implementation of complex scaling is discussed for a
DVR formulated in finite volume, one of the numerical techniques we use in this work to study CA-EC for three-body systems.

\subsubsection{Berggren basis expansion}
\label{Sec:Berggren}

\begin{figure}
    \centering
    \tikzsetnextfilename{contour}
\begin{tikzpicture}

  \draw[thin,gray,step=0.5,dotted] (-1,-2.5) grid (5,2.5);
  \draw[->, thick] (-1,0) -- (5,0) node[right] {$\Rp \, k$};
  \draw[->, thick] (0,-2.5) -- (0,2.5) node[above] {$\Ip \, k$};

  \draw[domain=0:5,samples=100,color=red!50,thick] plot (\x, -0.5 *\x * exp{-0.03 * \x^4});
  \foreach \x/\y in {0/1, 0/0.6, 1/-0.2, 1.6/-0.3, 2/-0.18} \fill[red] (\x,\y) circle (1.5pt);

  \draw[domain=0:5,samples=10,color=blue!50,thick] plot (\x, -0.5 * \x);

\end{tikzpicture}
    \caption{Illustration of a typical Berggren basis (red) comparing the shape of the contour to the complex-scaling method (blue). Unlike the Berggren basis, the complex-scaled basis does not contain discrete states.}
    \label{fig:contour}
\end{figure}
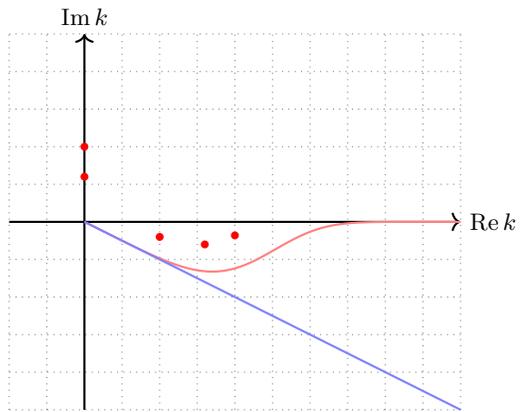

The Berggren basis~\cite{Berggren:1968zz,Berggren:1993zz} is defined by a single-particle completeness relation that includes discrete states (\ie, bound states and decaying resonances) and a continuum of scattering states. 
For a given partial wave $c=(l,j)$ with angular momentum $l$ and total spin $j$, it is defined in the momentum plane as:
\begin{equation}
	\sum_{i} \ket{ {u}_{c} ( {k}_{i} ) } \bra{ \tilde{u}_{c} ( {k}_{i} ) } + \int_{ \mathcal{L}_{c}^{+} } dk \, \ket{ {u}_{c} (k) } \bra{ \tilde{u}_{c} (k) } = \hat{\mathbf{1}}_{c} \,.
\label{eq_BB}
\end{equation}
Here, the sum goes over selected resonant states $\ket{ {u}_{c} ( {k}_{i} ) }$, or poles of the single-particle scattering matrix, defined by their momenta $k_i$, and the integral goes over a continuum of complex-momenta scattering states $\ket{ {u}_{c} (k) }$ along a contour $\mathcal{L}_c^+$ surrounding the selected discrete states in the fourth quadrant before going to $k \to \infty$. 
The tildes in Eq.~\eqref{eq_BB} represent time reversal, and $\hat{\mathbf{1}}_{c}$ denotes the identity operator in the $c$ partial-wave space~\cite{de_la_madrid_resonance_2005}.
An illustration of a typical Berggren basis in contrast to a simple complex-scaled basis is shown in Fig.~\ref{fig:contour}.

In practice, the Berggren basis can be generated for each partial wave using selected eigenstates of any single-particle Hamiltonian suitable for the problem at hand. 
The selection of the relevant discrete states is determined by the physics one wishes to study.
For completeness of the Berggren basis to hold, one needs to include the discrete states corresponding to the poles that are located above the chosen contour, as shown in Fig.~\ref{fig:contour}.
However, it is \emph{not} necessary to choose a contour that encompasses all poles that a given choice of potential might generate.
For example, if one is interested in describing the narrow low-lying resonances of the system considered, it is generally not necessary to define a contour going deep into the fourth quadrant to include broad resonances.
It is also possible to not use any potential, except for the centrifugal barrier, in which case the Berggren basis reduces to the integral entering Eq.~\eqref{eq_BB} and the basis states are spherical Bessel functions, possibly continued in the complex plane.
In that case, and in the limit where the contour is just a straight line into the fourth quadrant, one recovers the uniform complex scaling.

Due to its versatility, the Berggren basis is particularly well suited for many-body applications because continuum states can be included only in the partial waves known to be relevant for the problem at hand,\footnote{If little or nothing is known about a particular system of interest, one would still construct a basis that allows for all continuum states that \emph{might} be relevant.} and truncations can be applied selectively either at the single-particle level or at the many-body level.
Details about the Berggren basis can be found in Refs.~\cite{michel09_2,michel21_b260} in the context of the Gamow shell model.

\subsection{CA-EC with complex scaling for few/many-body Gamow states}
\label{sec:CA-EC}

The original EC method was introduced to efficiently approximate the eigenvectors of a Hermitian Hamiltonian $H(c)$, depending smoothly on the set of parameters $c$, using a small number of known exact eigenstates for different values of $c$ called ``training points.''
This is achieved by projecting the Hamiltonian $H(c_*)$ for the target set of parameters $c_*$ of interest onto the reduced space spanned by the training states.
Because these training states are in general not orthogonal, one then solves the generalized eigenvalue problem shown in Eq.~\eqref{eq:EC-GEVP}:
\begin{equation}
 H_{\text{EC}} \ket{\psi(c_*)}_{\text{EC}}
 = E(c_*)_{\text{EC}} \, N_{\text{EC}} \ket{\psi(c_*)}_{\text{EC}} \,.
\label{eq:EC-GEVP}
\end{equation}
The projected Hamiltonian $H_{\text{EC}}$ and norm matrix $N_{\text{EC}}$ are defined as
\begin{align}
\label{eq:EC-H}
 \big(H_{\text{EC}}\big)_{ij} &= \braket{\psi(c_i)|H(c_*)|\psi(c_j)} \,, \\
\label{eq:EC-N}
 \big(N_{\text{EC}}\big)_{ij} &= \braket{\psi(c_i) | \psi(c_j)} \,.
\end{align}
In Ref.~\cite{yapa_eigenvector_2023}, we introduced the CA-EC method to perform bound-to-resonance (B2R) extrapolations in two-body systems.
This was achieved by generalizing the EC method to non-Hermitian Hamiltonians generated with the uniform complex-scaling method.
The matrix elements in Eqs.~\eqref{eq:EC-H} and~\eqref{eq:EC-N} are then evaluated
using the so-called ``c-product''~\cite{Moiseyev:1978aa,Moiseyev:2011}, which, compared to
the standard inner product, amounts to omitting complex conjugation for bra-side operands.
In this setup, bound states still have real energies, but their wave functions become genuinely complex, \ie, they cannot be made real by eliminating a global phase factor.
In CA-EC, the reduced basis is spanned by these complex bound-state wave functions, along with their complex conjugates.
Adding the latter is key to being able to obtain complex resonance energies in the reduced space.

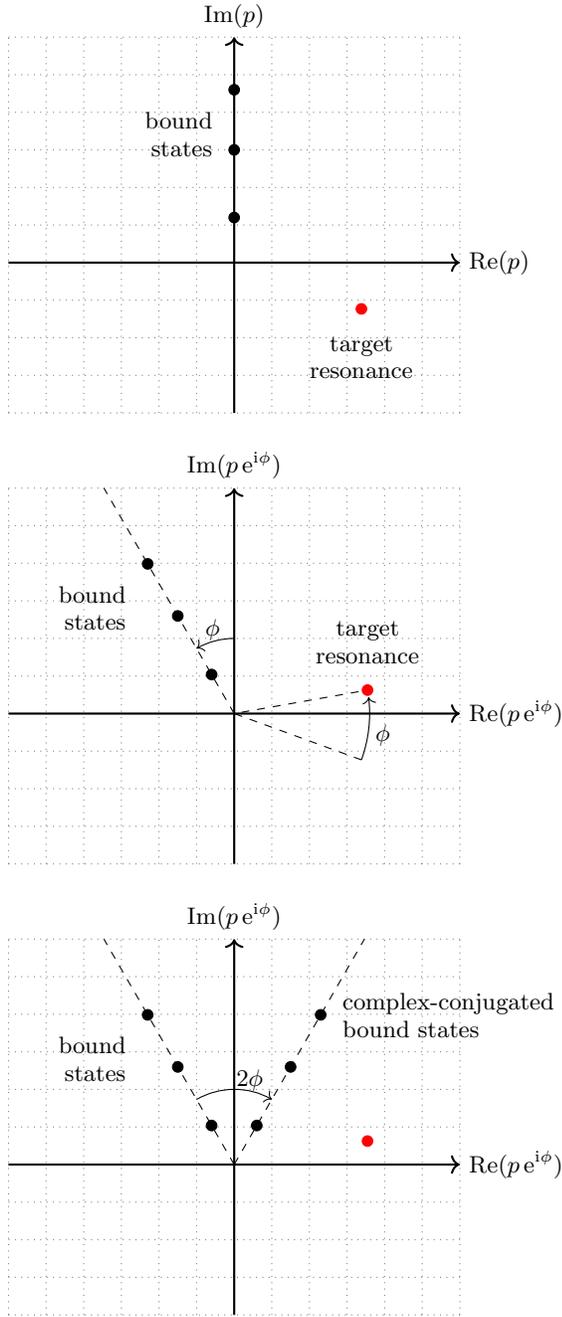
\begin{figure}
    \centering
    \tikzsetnextfilename{CAEC_illustration}
\begin{tikzpicture}

\def \targemagnitude {1.8}
\def \targetangle {-20}
\def \angle {30}
\def \boundks {0.6,1.5,2.3}

\begin{scope}[shift={(0,6)}]
  \draw[thin,gray,step=0.5,dotted] (-3,-2) grid (3,3);
  \draw[->, thick] (-3,0) -- (3,0) node[right] {$\Rp (p)$};
  \draw[->, thick] (0,-2) -- (0,3) node[above] {$\Ip (p)$};

  \draw[variable=\k, samples at=\boundks, mark=*, only marks] plot (0, \k) node[below left=5, align=right] {bound\\states};

  \draw[mark=*, mark options={red}] plot coordinates {({cos(\targetangle) * \targemagnitude}, {sin(\targetangle) * \targemagnitude})} node[below=7, align=center] {target\\resonance};
\end{scope}

\begin{scope}[shift={(0,0)}]
  \draw[thin,gray,step=0.5,dotted] (-3,-2) grid (3,3);
  \draw[->, thick] (-3,0) -- (3,0) node[right] {$\Rp (p \, \cphase{})$};
  \draw[->, thick] (0,-2) -- (0,3) node[above] {$\Ip (p \, \cphase{})$};

  \draw[variable=\k, samples at=\boundks, mark=*, only marks] plot ({sin(-\angle) * \k}, {cos(\angle) * \k}) node[below left=5, align=right] {bound\\states};
  \draw[variable=\x, domain=0:3, dashed] plot ({\x * tan(-\angle)}, \x);

  \draw[->] (0,1) arc (90 : {90 + \angle} : 1) node[above right] {$\phi$};

  \draw[dashed] ({cos(\targetangle) * \targemagnitude}, {sin(\targetangle) * \targemagnitude}) -- (0,0) -- ({cos(\targetangle + \angle) * \targemagnitude}, {sin(\targetangle + \angle) * \targemagnitude});
  \draw[->] ({cos(\targetangle) * \targemagnitude}, {sin(\targetangle) * \targemagnitude}) node[above right=2] {$\phi$} arc (\targetangle : {\targetangle + \angle - 3} : \targemagnitude);

  \draw[mark=*, mark options={red}] plot coordinates {({cos(\targetangle + \angle) * \targemagnitude}, {sin(\targetangle + \angle) * \targemagnitude})} node[above=7, align=center] {target\\resonance};
\end{scope}

\begin{scope}[shift={(0,-6)}]
  \draw[thin,gray,step=0.5,dotted] (-3,-2) grid (3,3);
  \draw[->, thick] (-3,0) -- (3,0) node[right] {$\Rp (p \, \cphase{})$};
  \draw[->, thick] (0,-2) -- (0,3) node[above] {$\Ip (p \, \cphase{})$};

  \draw[variable=\k, samples at=\boundks, mark=*, only marks] plot ({sin(-\angle) * \k}, {cos(\angle) * \k}) node[below left=5, align=right] {bound\\states};
  \draw[variable=\x, domain=0:3, dashed] plot ({\x * tan(-\angle)}, \x);

  \draw[variable=\k, samples at=\boundks, mark=*, only marks] plot ({sin(\angle) * \k}, {cos(\angle) * \k}) node[right=5, align=left] {complex-conjugated\\bound states};
  \draw[variable=\x, domain=0:3, dashed] plot ({\x * tan(\angle)}, \x);

  \draw[->] ({-sin(\angle)}, {cos(\angle)}) arc ({90 + \angle} : {90 - \angle} : 1) node[above left] {$2 \phi$};

  \draw[mark=*, mark options={red}] plot coordinates {({cos(\targetangle + \angle) * \targemagnitude}, {sin(\targetangle + \angle) * \targemagnitude})};
\end{scope}

\end{tikzpicture}
    \caption{Illustration of the CA-EC method. 
    \textbf{Top panel:} The asymptotic momenta of the training bound states originally lie on the positive imaginary axis, while that of resonances exist inside the fourth quadrant. 
    \textbf{Middle panel:} Complex scaling can be thought of as rotating the entire plane by an angle of $\phi$ counterclockwise. This effectively moves the resonances into the second quadrant, thereby making their wave functions convergent. 
    \textbf{Bottom panel:} Complex conjugation of the complex-scaled bound states, which effectively reflects their positions about the $y$ axis, places their asymptotic momenta in the second panel, in better proximity to the target resonance.}
    \label{fig:CAEC_illustration}
\end{figure}

The method can be intuitively understood in the two-body case. 
Indeed, in the CA-EC method with uniform complex-scaling, before the complex rotation, the asymptotic momenta\footnote{We use the expression "asymptotic momentum" to denote the complex value $p$ that determines the asymptotic wave function. It may not necessarily refer to the physical momentum, especially after complex conjugation.} associated with the training bound states lie on the positive imaginary axis, while the target (decaying) resonance lies in the fourth quadrant (see Fig.~\ref{fig:CAEC_illustration}).
After rotation by an angle $0 < \phi < 45\degree$, the now rotated bound states move into the second quadrant, while the target resonance moves into the first. 
This is where taking the complex conjugate of the rotated bound states is critical as it moves the asymptotic momenta of the rotated bound states into the first quadrant, together with the target resonance, giving them the same type of asymptotic behavior.

In fact, we can make the exact matching condition explicit in the two-body case.
Here, we provide the precise conditions on the training bound state energy and rotation angle that must be satisfied to make the asymptotic momentum of the complex-conjugated rotated bound state agree with the asymptotic momentum of the rotated resonance.
We denote the complex momenta of the target resonance and training bound state in polar coordinates as $\abs{k_R} \ee^{{-}\ii\theta_R}$ and $\abs{k_B} \ee^{\ii\pi/2}$, respectively, where $\theta_R$ is positive. 
After a clockwise rotation of the momentum plane by an angle $\phi$, in the new frame, the states will have the momenta $\abs{k_R} \ee^{\ii(\phi-\theta_R)}$ and $\abs{k_B} \ee^{\ii(\pi/2+\phi)}$, respectively. 
Taking the complex conjugate of the rotated bound state moves its asymptotic momentum to $\abs{k_B} \ee^{\ii(\pi/2-\phi)}$.
The matching condition is obtained by enforcing equality of the momenta
\begin{equation}
    \abs{k_R} \ee^{\ii(\phi-\theta_R)} = \abs{k_B} \ee^{\ii(\pi/2-\phi)} \,,
    \label{eq_match}
\end{equation}
which is satisfied when
\begin{align}
    & \abs{k_B} = \abs{k_R} \\
    & \phi = \frac{\pi}{4} + \frac{\theta_R}{2} \,.
    \label{eq_match_cond}
\end{align}
We see that if the resonance is narrow and lies near the real axis, $\theta_R \to 0$ and we must have $\phi \to \pi/4$. 
However, if the resonance is broad and lies near the ${-}45\text\textdegree$ line in the momentum plane, $\theta_R \to \pi/4$ and we must have $\phi \to 3\pi/8$ or $67.5\text\textdegree$.
We thus see that while an exact matching condition can be derived in this case, it cannot be achieved in practice as the rotation angle must satisfy $\phi < \pi/4$. 
This reaffirms that CA-EC needs, in general, at least more than one training point to work.

In the many-body case, we argue that essentially the same happens, in the sense that taking the complex conjugate of training bound states expressed in a complex basis gives new training states with the same type of asymptotic as that of the target resonance we wish to extrapolate to.
This can be seen as follows:
if we restrict ourselves to a potential with a strict finite range $R$, \ie,
\begin{equation}
    V(\vec x_1, \vec x_2, \ldots) = 0 \quad \text{for} \quad | \vec x_i - \vec x_j | > R \quad \forall \ i,j \, ,
\end{equation}
we see that a bound state $\psi(\vec r_1, \vec r_2, \ldots)$ satisfies
\begin{spliteq}
    H \, \psi(\vec r_1, \vec r_2, \ldots) &= -\sum_{i=1}^{n-1} \frac{\nabla_i^2}{2 \mu_i} \, \psi(\vec r_1, \vec r_2, \ldots) \\
    &= E \, \psi(\vec r_1, \vec r_2, \ldots) \quad \text{as} \quad |\vec r_i| \rightarrow \infty \,.
\end{spliteq}
In a complex-scaled basis (indicated by $\phi$ in subscript) we have, similarly, 
\begin{spliteq}
    H_\phi \, \psi_\phi(\vec r_1, \vec r_2, \ldots) &= -\cphase{-2} \sum_{i=1}^{n-1} \frac{\nabla_i^2}{2 \mu_i} \, \psi_\phi(\vec r_1, \vec r_2, \ldots) \\
    &= E \, \psi_\phi(\vec r_1, \vec r_2, \ldots) \quad \text{as} \quad |\vec r_i| \rightarrow \infty \,,
\end{spliteq}
and taking the complex conjugate of the above results in
\begin{multline}
    {-}\cphase{2} \sum_{i=1}^{n-1} \frac{\nabla_i^2}{2 \mu_i} \, \psi^*_\phi(\vec r_1, \vec r_2, \ldots) \\
    = E \, \psi^*_\phi(\vec r_1, \vec r_2, \ldots) \quad \text{as} \quad |\vec r_i| \rightarrow \infty
    \label{eq:ManyBodyCC}
\end{multline}
because $E$ corresponds to a bound state and is therefore real. 
Multiplying both sides by $\cphase{-4}$ gives
\begin{multline}
    {-}\cphase{-2} \sum_{i=1}^{n-1} \frac{\nabla_i^2}{2 \mu_i} \, \psi^*_\phi(\vec r_1, \vec r_2, \ldots) \\
    = \cphase{-4} E \, \psi^*_\phi(\vec r_1, \vec r_2, \ldots) \quad \text{as} \quad |\vec r_i| \rightarrow \infty \,,
\end{multline}
or 
\begin{multline}
    H_\phi \, \psi^*_\phi(\vec r_1, \vec r_2, \ldots) \\
    = \cphase{-4} E \, \psi^*_\phi(\vec r_1, \vec r_2, \ldots)
    \quad \text{as} \quad |\vec r_i| \rightarrow \infty \,.
\end{multline}

Overall, we have found that in the asymptotic region the complex-conjugated rotated bound state $\psi^*_\phi$ behaves like an eigenstate with a energy eigenvalue $\cphase{-4} E$. 
The angle $\phi$ can be chosen so that the eigenvalue falls close to the energy of the target resonance state.
In practice, CA-EC uses multiple training bound states for a fixed complex-scaling angle $\phi$, which together provide sufficient flexibility to describe resonances with different energies.

\subsection{CA-EC in the Berggren basis}
\label{Sec:result2bGSM}

The arguments above are based on the uniform complex-scaling method and do not immediately generalize to the Berggren basis. 
In the complex-scaling method, the training bound states are expressed using states along a straight path in the fourth quadrant of the momentum plane, which can be defined by a single complex phase, but in the Berggren basis, they are expressed using states on a deformed contour in the same quadrant, in addition to, possibly, a few resonant states also in the fourth quadrant or on the positive imaginary axis. 
However, from the point of view of the CA-EC method, what matters is to have training states that can help reproduce the outgoing asymptotic of the target resonance state.
In that regard, the training bound states will contain information about decay in both the Berggren basis and the complex-scaling method, and this information can be ``extracted'' via the use of the complex conjugation.

As noted above, the Berggren basis may contain discrete states (poles) in addition to the continuum states along the contour.
These discrete states are absent in the complex-scaling method but are usually important in Berggren basis applications to the extent that they help describe the inner (localized) part of the wave function. 

In the CA-EC method in the Berggren basis, because single-particle pole states all have specific asymptotics which may or may not contribute to that of the many-body target state, we expect their contribution to the target asymptotic to be comparable to that of any randomly selected scattering state. 

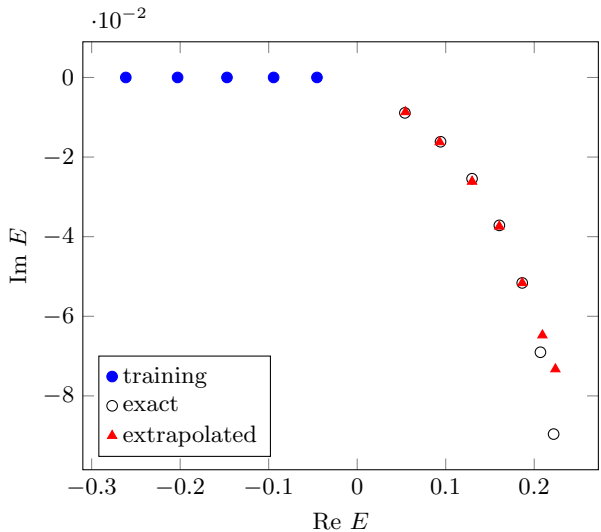
\begin{figure}
    \centering
    \tikzsetnextfilename{2b_GSM_B2R}
\begin{tikzpicture}
\begin{axis}[
    xlabel={$\Rp~E$},
    ylabel={$\Ip~E$},
    legend pos=south west,
    legend cell align={left},
    legend entries={training, exact, extrapolated},
    yticklabel style={/pgfplots/scaled ticks=false, /pgf/number format/fixed, /pgf/number format/precision=10}
]
\addplot [
    scatter,
    only marks,
    point meta=explicit symbolic,
    scatter/classes={
        training={blue},
        exact={mark=o,draw=black},
        extrapolated={mark=triangle*,red}
    },
] table [meta=label] {data/2b_GSM_B2R.csv};
\end{axis}
\end{tikzpicture}
    \caption{Berggren basis CA-EC for a two-boson system.}
    \label{fig:2b_GSM_B2R}
\end{figure}

To demonstrate that CA-EC can work in the Berggren basis, we consider a two-body toy model with the following potential:
\begin{equation}
    V(r) = c \left[ {-}5 \, \ee^{{-}\frac{r^2}{3}} + 2 \, \ee^{{-}\frac{r^2}{10}} \right] \,.
\end{equation}
We build a Berggren basis using this potential for $c=0.6$ in the $S$ wave, with a contour consisting of three segments defined by the points $k_0 = 0.0$, $k_1 = 0.4 - 0.15\ii$, $k_2 = 0.8$ and $k_3 = 0.6$, and a resonant pole at $E \approx 0.168 - 0.041 \ii$. The contour was discretized according the Gauss-Legendre quadrature with 128 points per segment.
We then use this Berggren basis to generate five training bound states by changing $c$, and finally proceed to a CA-EC B2R extrapolation.
As can be seen in Fig.~\ref{fig:2b_GSM_B2R}, the extrapolation remains very accurate until it reaches the resonances with the largest widths.
This shortfall can be mitigated by enlarging the contour in the Berggren basis to handle broader resonances.
Overall, although the mathematical explanation why CA-EC strictly speaking pertains to uniform complex scaling (straight line contour), we find that CA-EC also performs well for B2R extrapolations with the Berggren basis, which is relevant due to the scalability of this method to larger systems (discussed in detail in Ref.~\cite{michel09_2}).
Since it is possible to straighten and elongate the left portion of a Berggren-basis contour,
we argue that the good performance of CA-EC can be understood by considering this limit.

\subsection{Extrapolating from narrow to broad resonances}
\label{Sec:2b_GSM_R2R}

Bound-to-resonance CA-EC extrapolations have the advantage of relying only on bound-state training eigenvectors, which are typically easier to obtain than resonances. 
However, current many-body methods formulated in the quasistationary formalism are often reliable when it comes to calculating narrow many-body resonances. 
For that reason, here we show empirically an intriguing property of EC extrapolations (without using CA-EC) in the Berggren basis going from narrow resonances to broad resonances. 

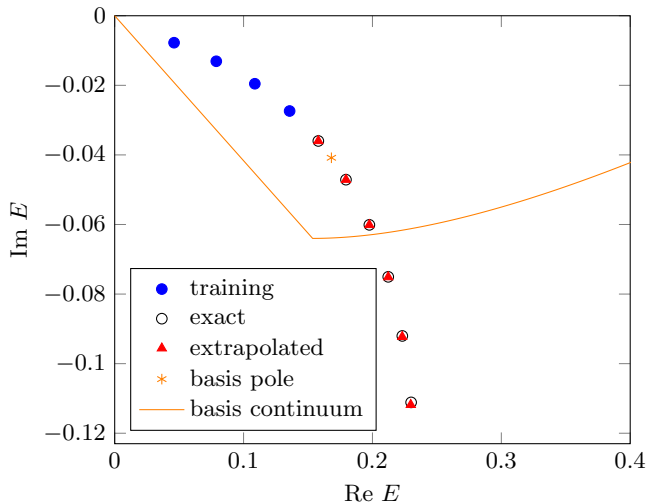
\begin{figure}
    \centering
    \tikzsetnextfilename{2b_GSM_R2R}
\begin{tikzpicture}
\begin{axis}[
    xmin=0,
    xmax=0.4,
    ymax=0,
    xlabel={$\Rp~E$},
    ylabel={$\Ip~E$},
    legend pos=south west,
    legend cell align={left},
    legend entries={training, exact, extrapolated, basis pole, basis continuum},
    yticklabel style={/pgfplots/scaled ticks=false, /pgf/number format/fixed, /pgf/number format/precision=10}
]

\addplot [
    scatter, only marks,
    point meta=explicit symbolic,
    scatter/classes={
        training={blue},
        exact={mark=o, draw=black},
        extrapolated={mark=triangle*, red},
        basis={mark=asterisk, orange}
    }
] table [meta=label] {data/2b_GSM_R2R.csv};

\addplot [scatter, no marks, orange] table {data/2b_GSM_R2R_contour.csv};

\end{axis}
\end{tikzpicture}
    \caption{Narrow-to-broad EC for a two-boson system in Berggren basis.}
    \label{fig:2b_GSM_R2R}
\end{figure}

Using the setup discussed in Sec.~\ref{Sec:result2bGSM}, we generated a set of narrow resonances as training points, with energies enclosed by the contour of the Berggren basis used to expressed their wave functions, as shown in Fig.~\ref{fig:2b_GSM_R2R} (blue points).
We note that the Berggren basis considered also contains one resonant pole. 
We then perform EC extrapolations of resonances with energies falling \emph{outside} the contour and find that these agree remarkably well with exact calculations for these parameters (performed using a modified contour that properly encloses these poles).
This result suggests that, in a sense, the EC basis outperforms the original Berggren basis, considering the latter should not be able to accurately describe states outside the contour~\cite{Berggren:1968zz}. 
While this might seem surprising, the EC matrix elements defined in Eqs.~\eqref{eq:EC-H} and~\eqref{eq:EC-N} are, in principle, independent of the underlying common basis in which the training eigenvectors are calculated, and thus we should just see the training eigenvectors as a new (reduced) basis in which the non-Hermitian problem is being solved.
However, it is obvious that the EC basis cannot ``outspan'' the original basis as the eigenvectors of the EC Hamiltonians can all be expressed in terms of the original basis,
as linear combinations.
The reason why the EC extrapolation appears to outperform the original basis is likely that the diagonalization of the Hamiltonian expressed in the original basis cannot converge onto the desired eigenstate due to the underlying lack of completeness of the Berggren basis for the target state. 
In contrast, the diagonalization of the reduced EC Hamiltonian is performed in a space spanned by true eigenvectors within a limited region of the parameter space. 
This effectively constraints the structure of EC solutions, most likely the inner part of their wave functions, so that their associated eigenvalues remain on a smooth trajectory.
That is, while the eigenstates of the reduced Hamiltonian may not be actual eigenstates of the full Hamiltonian when expanded back into the original underlying basis, they can still yield accurate approximations of the physical resonance energies in the reduced space.
This finding could prove useful to significantly improve the reach and accuracy of EC extrapolations in large-scale many-body problems where narrow resonances can be calculated.

\section{Applications to three-body systems}
\label{sec:Results}

In Sec.~\ref{sec:3boson}, we start by showing that CA-EC can perform B2R extrapolations for genuine three-body resonances using the complex-scaled CI method.
Specifically, we use a HO basis in Jacobi coordinates to describe the three-body system, and benchmark the results against the complex-scaled DVR approach of Ref.~\cite{yu_complex_2024}, which uses a different basis to describe few-body states in a periodic cubic volume.
Then, we replace the uniform complex-scaling method by a Berggren basis expansion and show that CA-EC continues to perform well and at a comparable computational cost.
Finally, we turn to fermionic systems and provide a realistic application based on the Gamow shell model in Sec.~\ref{sec:GSM}.

\subsection{Three-boson benchmark calculations}
\label{sec:3boson}

\paragraph*{Harmonic oscillator basis with complex scaling.}
First, we study bound-to-resonance extrapolation using an HO basis calculation formulated with Jacobi coordinates.

In this approach, we represent the Hamiltonian in a truncated space of
HO basis functions, which are given by~\cite{heyde_nuclear_1994}
\begin{multline}
    \braket{r,\theta,\phi|nlm} = \sqrt{\frac{\omega^{l+3/2}}{\sqrt\pi} \frac{2^{n+l+2} n!}{(2n+2l+1)!!}} \\ r^{l+1}
    \exp\left(-\frac{\omega r^2}{2}\right) L_n^{\left(l+\frac{1}{2}\right)}(\omega r^2) \, Y_l^m(\theta,\phi) \, ,
\label{eq:HO_basis}
\end{multline}
where $L_n^\alpha(x)$ are the generalized Laguerre polynomials, $n=0,1,\ldots$ is the radial quantum number that enumerates the $l$th partial wave basis, for which $m$ labels the angular momentum projection.
These can be used to derive the kinetic energy matrix elements in closed form~\cite{binder_effective_2016},
\begin{multline}
    \braket{n' l' m' | H_0 | n l m} = \frac{\omega}{2} \Biggl[ \delta_{n,n'} \left( 2 n + l + \frac{3}{2} \right) - \\
    \delta_{1,|n-n'|} \sqrt{n_> \left( n_> + l + \frac{1}{2} \right)} \Biggl] \delta_{l,l'} \delta_{m,m'} \, ,
\end{multline}
with $n_>=\max\{n,n'\}$, whereas potential energy matrix elements for a spherically symmetric
potential $V$ are calculated as
\begin{equation}
    \braket{n' l' m' | V | n l m} = \braket{n'|V|n} \delta_{l,l'} \delta_{m,m'} \, .
\end{equation}

It can be easily shown via Eq.~\eqref{eq:HO_basis}, that complex scaling in the HO basis can be trivially implemented by modifying the oscillator frequency as 
\begin{equation}
    \omega \to \omega \, \cphase{-2} \, ,
\end{equation}
where we have chosen $\phi = 0.1$ for this calculation.

The three-body basis states $\ket{n_1 l_1 n_2 l_2 ; L m_L}$ consist of linear combinations of tensor products of the states in Eq.\eqref{eq:HO_basis} for two Jacobi coordinates $r_1$ and $r_2$, corresponding to quantum numbers $(n_1, l_1)$ and $(n_2, l_2)$, respectively, such that the total angular momentum is fixed to some $L$ and $m_L$.\footnote{The matrix elements, and therefore the resulting spectrum, should be independent of $m_L$. Therefore we may always assume $m_L=0$.}
In this particular calculation, we focus on an $L=0$ state.
We use a basis truncation scheme where $2 n_1 + l_1 + 2 n_2 + l_2 \leq 40$.

While the construction of the kinetic energy part remains trivial, evaluation of the pairwise interactions requires the Moshinsky transformation~\cite{buck_simple_1996, moshinsky_transformation_1959}
corresponding to the coordinate transformation
\begin{equation}
    \begin{pmatrix}
    \vec r_{32} \\
    \vec r_{31}
    \end{pmatrix}
    =
    \begin{pmatrix}
    \sfrac{1}{2} & -1\\
    \sfrac{1}{2} & 1
    \end{pmatrix}
    \begin{pmatrix}
    \vec r_1\\
    \vec r_2
    \end{pmatrix} \, ,
\label{eq:Moshinsky}
\end{equation}
where $\vec r_{32}, \vec r_{32}$ are the usual relative vectors.
We utilize the numerical framework presented in Refs.~\cite{efros_calculation_2021, efros_oscillator_2023} to efficiently construct this transformation.

\begin{figure}
    \tikzsetnextfilename{3b-B2R}
\begin{tikzpicture}
    \begin{groupplot}[
        group style={group size=1 by 3, vertical sep=0pt, xticklabels at=edge bottom},
        height=5cm,
        width=\axisdefaultwidth,
        ylabel={$\Ip~E$},
        xmin=-2.7,
        xmax=4.8,
        ymin=-0.017,
        ymax=0.001,
        scaled y ticks=false,
        yticklabel style={/pgf/number format/fixed, /pgf/number format/precision=10}
    ]

    \nextgroupplot    
    \addplot [
        scatter,
        only marks,
        point meta=explicit symbolic,
        scatter/classes={
            training={blue},
            exact={mark=o,draw=black},
            extrapolated={mark=triangle*,red}
        }
    ] table [meta=label] {data/HO_B2R.csv};

    \node[above=2.8cm,right=6cm] {(a)};

    \nextgroupplot
    \addplot [
        scatter,
        only marks,
        point meta=explicit symbolic,
        scatter/classes={
            training={blue},
            exact={mark=o,draw=black},
            extrapolated={mark=triangle*,red}
        }
    ] table [meta=label] {data/FV.csv};

    \node[above=2.8cm,right=6cm] {(b)};
    
    \nextgroupplot[
        legend entries={training, exact, extrapolated},
        legend pos=south west,
        legend cell align={left},
        xlabel={$\Rp~E$}]
    \addplot [
        scatter,
        only marks,
        point meta=explicit symbolic,
        scatter/classes={
            training={blue},
            exact={mark=o,draw=black},
            extrapolated={mark=triangle*,red}
        }
    ] table [meta=label] {data/Berggren_B2R.csv};

    \node[above=2.8cm,right=6cm] {(c)};

    \end{groupplot}
\end{tikzpicture}
    \caption{Bound-to-resonance extrapolation for the three-boson system carried out under (a) complex-scaled CI in HO basis, (b) complex-scaled finite-volume basis and (c) Berggren basis.}
    \label{fig:3b-B2R}
\end{figure}
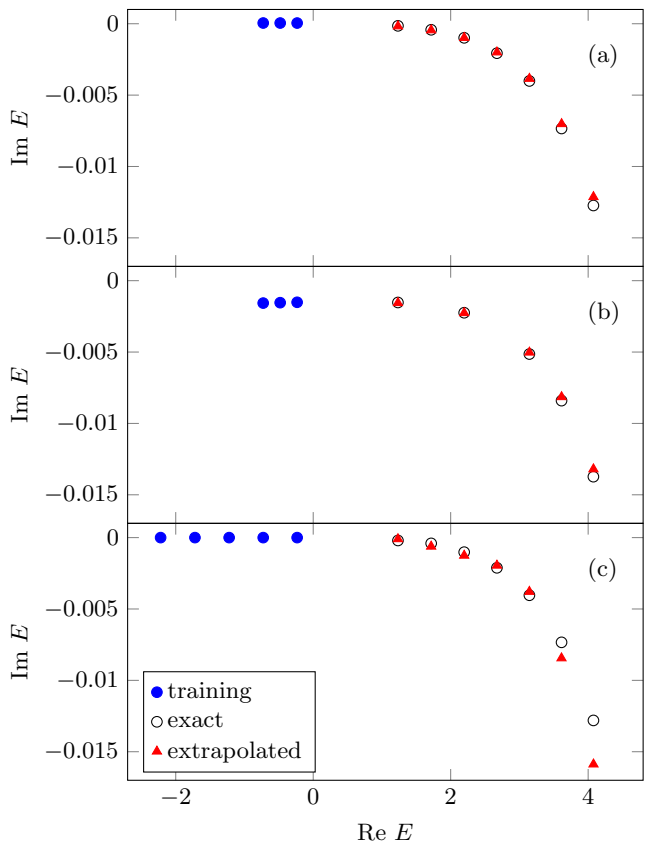

We consider for this test a toy model that includes pairwise two-body interactions between the particles, given by the potential
\begin{equation}
    V(r) = 2 \, \ee^{{-}\left(\frac{r-3}{1.5}\right)^2} \,.
\end{equation}
It was shown in Refs.~\cite{Klos:2018sen,yu_complex_2024} that this system supports a bosonic three-body resonance near $\Rp \, E \approx 4.1$.
In the finite-volume calculation of Ref.~\cite{Klos:2018sen} this state was identified in the $A_1^+$ representation of the cubic group, so one should expect that in infinite volume the system supports an $S$-wave (total angular momentum $L=0$) resonance.
Moreover, by means of adding a short-range three-body force, Ref.~\cite{Klos:2018sen} identified this state to be a genuine three-body resonance (without bound two-body decay channels).
This system can be made artificially bound by adding an attractive term so that
\begin{equation}
    V(r) = 2 \, \ee^{-\left(\frac{r-3}{1.5}\right)^2} - c \, \ee^{{-}\left(\frac{r}{3}\right)^2} \,,
\end{equation}
where $c$ is a prefactor controlling the degree of the added attraction.
We then perform B2R extrapolations by varying $c$ to obtain the results shown in Fig.~\ref{fig:3b-B2R}(a).

\paragraph*{Finite volume DVR calculation.} 
To demonstrate that CA-EC works independent of the particular few-body method used to study the system, we now use a finite-volume (FV) construction with periodic boundary conditions and relative coordinates in the discrete variable representation (DVR).
The calculation is shown in Fig.~\ref{fig:3b-B2R}(b).

The FV-DVR basis can be concisely introduced based on the set of following one-dimensional (1D) basis functions for a two-body system with relative coordinate $r$,
\begin{equation}
 \braket{r|k} = \frac{1}{\sqrt{NL}} \sum_{j=-N/2}^{N/2-1} \exp\left[\ii 2 \pi j \left(\frac{r}{L} - \frac{k}{N}\right) \right] \,,
\label{eq:Kronecker}
\end{equation}
for $k = {-}N/2, \ldots ,N/2-1$, where $N \in 2\mathbb{N}^+$ is to be understood as the number of lattice sites in one direction that partitions the distance $L$ into equidistant steps.
To generalize this approach to $d>1$ dimensions and $n$ particles, one works in the simple relative coordinates $\tilde{\vec{r}}_i$ and starts from
basis states as in Eq.~\eqref{eq:Kronecker} for each coordinate component, 
as detailed in Refs.~\cite{Klos:2018sen,Konig:2022cya}.

The advantages of the DVR becomes apparent when one considers the $N \rightarrow \infty$ limit, where the basis functions approach the Dirac $\delta$ function
\begin{equation}
    \braket{r|k} \approx \delta \left( \frac{r}{L} - \frac{k}{N} \right) \, ,
\end{equation}
which allows the potential energy for a local potential $V$ to be conveniently approximated as a diagonal matrix. 
Moreover, the 1D momentum operator matrix element takes the following closed form:
\begin{equation}
 \bra{k} p \ket{l} = \frac{\pi (-1)^{k-l}}{\ii L}
  \begin{cases}
    {-}\ii & \text{if } k=l\\
    \frac{\exp \left[ {-}\ii\frac{\pi (k-l)}{N} \right]}
    {\sin{\frac{\pi (k-l)}{N}}}
    & \text{otherwise}
  \end{cases} \, .
\label{eq:q-braket}
\end{equation}
This can be used to construct the overall kinetic-energy operator for an $n$-body system.
More details about the FV-DVR and in particular its efficient implementation for few-body states can be found in Refs.~\cite{Klos:2018sen,Konig:2022cya}.
Those references also explain how to implement symmetrization (or antisymmetrization) and projection onto definite parity subspaces for the DVR basis.
Implementation of complex scaling for the FV-DVR then proceeds as described in Ref.~\cite{yu_complex_2024}, following the procedure described in Sec.~\ref{Sec:cxscaling}.

Note that finite-volume effects induce some small discrepancies in the calculation. 
This is exemplified by bound states that do not lie exactly on the $\Rp~E=0$ line. 
For two-body systems, it is possible to extrapolate results to infinite volume because the volume dependence is known~\cite{yu_complex_2024}.
For the three-body system we study here, the effect is mitigated by taking a large
value for the box size $L$.
Specifically, the results shown in Fig.~\ref{fig:3b-B2R}(b) were obtained with
$L=15$, $N=24$ for the DVR basis, and the system was explicitly projected into a subspace of three-boson states with positive parity.

\paragraph*{Berggren basis.} 
Finally, we also repeat the calculation using a Berggren basis generated by a free Hamiltonian (\ie, without discrete pole states in the basis). 
The contour consists of three segments with the vertices at $k_0 = 0$, $k_1 = 4 \ee^{-0.1\ii}$, $k_2 = 5$, and $k_3 = 6$, discretized into $40$, $12$ and $15$ points respectively. Only $l=0,\ldots,6$ partial waves were included. The results are shown in Fig.~\ref{fig:3b-B2R}(c).

While the Moshinsky transform is directly applicable in the HO basis, a Berggren basis implementation requires an additional step.
One may calculate the potential energy matrices corresponding to $V(r_{32}) + V(r_{31})$ within an HO basis, and then use the appropriate unitary transformation to bring it back to the Berrgren basis. 
$V(r_{21})$, on the other hand, can be calculated directly on the Berggren basis without involving such transformations.

The results obtained with the Berggren basis are shown in panel (c) of Fig.~\ref{fig:3b-B2R}. 
We see that all three methods give qualitatively comparable results. 

\subsection{Fermionic systems: Gamow shell model application in $^6$He}
\label{sec:GSM}

Having demonstrated the feasibility of B2R extrapolations of genuine three-body bosonic resonances using CA-EC, for two different approaches using complex scaling as well as in the Berggren basis, we now turn our attention towards the fermionic case and provide a first realistic application of CA-EC by applying it to the \isotope[6]{He} isotope.
In nature, the ground state of \isotope[6]{He} is a halo state with peculiar effective three-body dynamics involving a tightly bound core of \isotope[4]{He} and two weakly bound neutrons, in which no two-body subsystem (\isotope[5]{He} or $2n$) is bound, making it a so-called Borromean state.
Rendering this state artificially unbound by reducing the strength of the interaction between the neutrons is thus a convenient way to obtain a genuine fermionic three-body resonance. 

To study this somewhat realistic situation, we design a simple model based on Ref.~\cite{fossez18_2171}, in which the core-nucleon interaction is represented by a Woods-Saxon potential reproducing the \isotope[4]{He}-$n$ phase shifts up to about 20.0 MeV, and the $n$-$n$ interaction is represented by a contact interaction in the $(S,T)=(0,1)$ channel controlled by a single parameter, adjusted as needed.
To solve the quantum many-body problem, we use the Gamow shell model code published in Ref.~\cite{michel21_b260}, which is a generalization of the shell model in the complex-energy plane using the Berggren basis. 
Within this approach, we consider a model space comprised of the $p_{3/2}$, $p_{1/2}$, $d_{5/2}$, and $d_{3/2}$ partial waves.
While the $d$ waves are represented with a single HO shell, the $p$ waves are represented in the Berggren basis. 
In the latter case, we include one discrete state set at $k=0.152\ii$ and $k=0.218-0.058\ii$ for the $0p_{3/2}$ and $0p_{1/2}$ shells, respectively, and a contour defined by the momenta $k_0=0.0$, $k_1=0.2-0.2\ii$, $k_3=1.0$, and $k_4=4.0$ (all in fm$^{{-}1}$). 
We note that the first segment of the contour goes down into the fourth quadrant of the momentum plane with an angle of ${-}\pi/4$.

\begin{figure}
    \centering
    \tikzsetnextfilename{6He_GSM_CAEC}
\begin{tikzpicture}
    \begin{groupplot}[
        group style={group size=1 by 3, vertical sep=0pt, xticklabels at=edge bottom},
        height=5cm,
        width=\axisdefaultwidth,
        ylabel={$\Ip~E$ (MeV)},
        xmin=-3.0,
        xmax=2.0,
        ymin=-0.40,
        ymax=0.1,
        scaled y ticks=false,
        yticklabel style={/pgf/number format/fixed, /pgf/number format/precision=10}
    ]

    \nextgroupplot    
    \addplot [
        scatter,
        only marks,
        point meta=explicit symbolic,
        scatter/classes={
            training={blue},
            exact={mark=o,draw=black},
            extrapolated={mark=triangle*,red}
        }
    ] table [meta=label] {data/data_6He_EC_10.csv};

    \node[above=2.8cm,right=6cm] {(a)};

    \nextgroupplot    
    \addplot [
        scatter,
        only marks,
        point meta=explicit symbolic,
        scatter/classes={
            training={blue},
            exact={mark=o,draw=black},
            extrapolated={mark=triangle*,red}
        }
    ] table [meta=label] {data/data_6He_CAEC_3.csv};

    \node[above=2.8cm,right=6cm] {(b)};

    \nextgroupplot[
        legend entries={training, exact, extrapolated},
        legend pos=south west,
        legend cell align={left},
        xlabel={$\Rp~E$ (MeV)}]

    \addplot [
        scatter,
        only marks,
        point meta=explicit symbolic,
        scatter/classes={
            training={blue},
            exact={mark=o,draw=black},
            extrapolated={mark=triangle*,red}
        }
    ] table [meta=label] {data/data_6He_CAEC_10.csv};

    \node[above=2.8cm,right=6cm] {(c)};

    \end{groupplot}
\end{tikzpicture}
    \caption{Bound-to-resonance extrapolations for the \isotope[6]{He} ground state energy using (a) the EC method with ten training states, (b) CA-EC with three training states, and (c) CA-EC with ten training states.}
    \label{fig:3b-B2R_6He}
\end{figure}
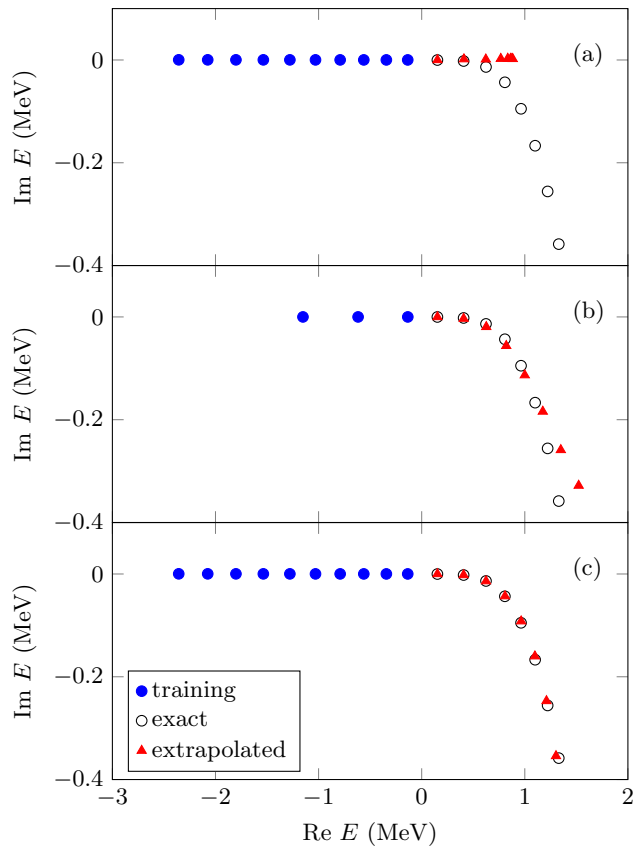

We first generate ten training bound states with energies between about ${-}2.3$ MeV and ${-}0.13$ MeV from uniformly spaced parameters, and then perform both EC and CA-EC extrapolations for states with energies going from about 0.15 MeV to 1.32 MeV, and with widths going from less than 1 keV up to 717 keV. 
The results, summarized in Fig.~\ref{fig:3b-B2R_6He}, show that the simple EC extrapolation [panel (a)] fails as expected, while the CA-EC extrapolation [panel (c)] closely matches the exact GSM results, with deviations below 10 keV for the widths.
To demonstrate how efficient CA-EC can be at extrapolating resonances, we also consider the same calculation with only three training bound states, with energies between about ${-}1.25$ MeV and ${-}0.13$ MeV. 
The range was chosen so that the most bound training state satisfies $\abs{E_B} \approx \abs{E_R}$, where $E_R$ is the energy of the penultimate resonance when going down the fourth quadrant. 
In this case, the CA-EC extrapolation [panel (b)] degrades somewhat, but still provides a fairly good estimate of the energy positions and widths of the targeted resonances.

This proof-of-principle in a realistic case demonstrates that accurate and scalable B2R extrapolations in many-body fermionic systems are possible using CA-EC in the Berggren basis. 

\subsection{Resonance involving a bound subsystem}
\label{sec:BoundSubsystem}

In this section, we conclude this work by considering a three-body system with a bound two-body subsystem. The first two particles interact with each other via the fixed interaction
\begin{equation}
 V_\text{bound}(r) = -2 \ee^{-\frac{r^2}{4}} \, ,
\end{equation}
which supports an $S$-wave bound subsystem at $E = -0.39196$.
The third particle interacts with the other two via the potential
\begin{equation}
    V_\text{res}(r) = c \left[ -\ee^{-\frac{r^2}{3}}+\ee^{-\frac{r^2}{10}} \right] \, .
\end{equation}
As before, the system transitions from bound to resonant states as the interaction is made weaker ($c$ is decreased). However, unlike before, the threshold is not located at $E=0$. Nevertheless, CA-EC works as expected, as shown in Fig.~\ref{fig:3b-subsystem}. This calculation was performed in a complex-scaled HO basis constructed identically to the one described in Sec.~\ref{sec:3boson}.
While we already established in previous work that CA-EC works well for genuine two-body
systems~\cite{yapa_eigenvector_2023}, the results presented here demonstrate that this remains true for a state with a two-body decay mode generated by genuine three-body dynamics.
\begin{figure}
    \centering
    \tikzsetnextfilename{3b-subsystem}
\begin{tikzpicture}
\begin{axis}[
    xlabel={$\Rp~E$},
    ylabel={$\Ip~E$},
    legend pos=south west,
    legend cell align={left},
    legend entries={training, exact, extrapolated},
    yticklabel style={/pgfplots/scaled ticks=false, /pgf/number format/fixed, /pgf/number format/precision=10}
]
\addplot [
    scatter,
    only marks,
    point meta=explicit symbolic,
    scatter/classes={
        training={blue},
        exact={mark=o,draw=black},
        extrapolated={mark=triangle*,red}
    },
] table [meta=label] {data/3b-subsystem.csv};
\end{axis}
\end{tikzpicture}
    \caption{Bound-to-resonance extrapolation for three-body system that decays into a two-particle bound subsystem and a spectator particle. Notice how the threshold is not located at $E=0$.}
    \label{fig:3b-subsystem}
\end{figure}
\section{Discussion and outlook}
\label{sec:Conclusion}

In this work, building upon Ref.~\cite{yapa_eigenvector_2023}, we extended the CA-EC method to perform efficient bound-to-resonance extrapolations for three-body systems and in particular for resonance states that exhibit genuine three-body decay. 
We showed that the CA-EC method can be formulated using not only the uniform complex-scaling method, but also with a Berggren basis expansion, thus opening the door to efficient CA-EC extrapolations in many-body systems. 
To demonstrate that CA-EC can be efficiently applied in realistic systems, we performed a CA-EC B2R extrapolation using the Berggren basis for the ground state of the \isotope[6]{He} isotope.

In the future, we will apply the CA-EC method in the many-body sector, and consider extensions towards anti-bound states to study systems at the limit of nuclear stability. 
At a more fundamental level, we will also investigate the precise relation between the uniform complex-scaling method and the Berggren basis expansion.

\begin{acknowledgments}
This work was supported in part by the National Science Foundation under Grants No.~PHY--2044632 and No.~PHY--2238752, and by the U.S. Department of Energy under the STREAMLINE Collaboration Awards No.~DE-SC0024520 and No.~DE-SC0024646. This material is based upon work supported by the U.S. Department of Energy, Office of Science, Office of Nuclear Physics, under the FRIB Theory Alliance Award No.~DE-SC0013617.
\end{acknowledgments}

\bibliographystyle{apsrev4-2}
\bibliography{refs}


\end{document}